\documentclass{JHEP3}

\usepackage{amsmath}
\usepackage{amssymb}
\usepackage{graphicx}

\newcommand{\AP}{{\alpha^{\prime}}}
\newcommand{\pd}{\partial}

\newcommand{\Fc}{\mathcal{F}}
\newcommand{\BDc}{\mathcal{D}}

\newcommand{\diag}{\mathop{\mathrm{diag}}\nolimits}

\title{Non-local SFT Tachyon and Cosmology}

\author{Alexey~S.~Koshelev\footnote{On leave from \textit{Steklov
Mathematical Institute of RAS, Gubkin st., 8, 119991, Moscow,
Russia, e-mail:} \texttt{koshelev@mi.ras.ru}}\\
Department of Physics, University of Crete, P.O. Box 2208, 71003,
Heraklion, Crete, Greece,\\
E-mail: \email{koshelev@physics.uoc.gr}}

\abstract{Cosmological scenarios built upon the generalized
non-local String Field Theory and $p$-adic tachyons are examined. A
general kinetic operator involving an infinite number of derivatives
is studied as well as arbitrary parameter $p$. The late time
dynamics of just the tachyon around the non-perturbative vacuum is
shown to leave the cosmology trivial. A late time behavior of the
tachyon and the scale factor of the FRW metric in the presence of
the cosmological constant or a perfect fluid with $w>-1$ is
constructed explicitly and a possibility of non-vanishing
oscillations of the total effective state parameter around the
phantom divide is proven.}


\preprint{\hepth{0701103}}

\begin{document}

\section{Introduction}

Contemporary cosmological observational data~\cite{data} strongly
support that the present Universe exhibits an accelerated expansion
providing thereby an evidence for a dominating Dark Energy (DE)
component. Recent results of WMAP~\cite{Spergel06} together with the
data on Ia supernovae give the following bounds for the DE state
parameter:
\begin{equation*}
w_{\text{DE}}=-0.97^{+0.07}_{-0.09}
\end{equation*}
or without an a priori assumption that the Universe is flat and
together with the data on large-scale structure and supernovae
$w_{\text{DE}}=-1.06^{+0.13}_{-0.08}$.

The phantom divide $w=-1$ separates the quintessence models,
$w>-1$~\cite{Wetterich,Peebles}, containing an extra light scalar
field which is not in the Standard Model set of fields~\cite{Okun},
the cosmological constant, $w=-1$~\cite{S-St,Padmanabhan-rev}, and
the ``phantom'' models, $w<-1$, which can be realized by a scalar
field with a ghost (phantom) kinetic term. In this case all natural
energy conditions are violated and there are problems of instability
both at the classical and quantum levels \cite{Caldwell,Woodard}.

Experimental data, as we see, do not contradict a possibility $w<-1$
and moreover the direct search strategy to test inequality $w<-1$
has been proposed~\cite{0312430}. Studying of such models attracts a
lot of attention. Some projects~\cite{Riess06} explore whether $w$
varies with the time or is an exact constant. Varying $w$ obviously
corresponds to a dynamical model of the DE which generally speaking
includes a scalar field. Modified models of General Relativity also
generate an effective scalar field (see for example~\cite{modGR} and
refs. therein). Other DE models based on brane-world scenarios are
presented in~\cite{brane}. An excellent review \cite{sami_review}
and references therein may provide the reader with a more detailed
discussion of the DE dynamics.

Models with a crossing of the $w=-1$ barrier are also a subject of
recent studies. Simplest ones include two scalar fields (one phantom
and one usual field, see~\cite{AKVtwofields,quint} and refs.
therein). General $\kappa$-essence models~\cite{mukhanov,wei} can
have both $w<-1$ and $w\geqslant -1$ but a dynamical transition
between these domains is forbidden under general
assumptions~\cite{Vikman} and is possible only under special
conditions~\cite{andrianov}.

In the present paper we investigate cosmological applications of a
scalar field model coming from the String Field Theory (SFT) tachyon
dynamics (see \cite{review-sft} for a review) described by the
action
\begin{equation*}
S=\frac1{g_o^2}\int d^dx\left(\frac{1}{2} \Phi\Fc(\Box)\Phi-
\frac{1}{p+1}\Phi^{p+1}(x)\right).
\end{equation*}
Here $g_o$ is a coupling constant. Function $\Fc(\Box)$ is not
specified explicitly. Such a theory describes in particular an
effective open SFT tachyon as well as a $p$-adic
formulation~\cite{BFOW,padic} of the tachyon dynamics. In both these
examples a kinetic operator $\Fc$ gives a non-local action. Analysis
of the Dirac-Born-Infeld approach to the tachyon cosmology may be
found in \cite{DBIreview,DBIcosm} and references therein.

The cosmological model incorporating the non-local dynamics of the
open SFT tachyon field proposed in \cite{Arefeva}. This model is
based on the SFT formulation of the fermionic NSR string with the
GSO$-$ sector \cite{NPB} and its cosmological applications were
studied in \cite{AKV,AK,AV}. A characteristic feature of this model
in the flat background is a presence of a rolling tachyon solution
\cite{AJK,yar}\footnote{Other works which address an issue of
solving the SFT equations of motion for homogenous time-dependent
tachyon profiles are, for instance, \cite{timetach} and refs.
therein.}. In the bosonic SFT, however, such a solution does not
exist \cite{Zw,VSV} at least in the flat space. The dynamics of a
non-local tachyon on a cosmological background in the
Hamilton-Jacobi formalism  is studied in \cite{calcagni}. It is
explicitly shown in \cite{AK} that the non-locality provides a
crossing of the $w=-1$ barrier in spite of the presence of only one
scalar field and being a string theory limit the model addresses all
stability issues \cite{SD} to the string theory\footnote{In
\cite{VR} it has been proposed a phantom model without an UV
pathology in which a vector field is used.}. In a recent paper
\cite{biswas} $p$-adic inflationary models are constructed using
similar non-local Lagrangian\footnote{See also \cite{biswas2} where
a non-local modification of gravity is studied.}.

Our present goal is a general analysis without specifying an exact
form of operator $\Fc$ and a qualitative analysis of linearized
Friedmann equations in this model at large times. From the point of
view of the SFT such a generalization is natural since form of the
kinetic operator depends on the contents of string excitations taken
into account as well as parameter $p$ will not be necessarily $3$ as
it is in the above cited papers. It is shown in the current paper
that the form of operator $\Fc$ is the most crucial ingredient. The
only physical assumption made is that a non-perturbative vacuum does
exist and there are no open string excitations in it since being
associated with the open string tachyon, our scalar field $\Phi$
describes according to Sen's conjecture \cite{Sen-g} a transition of
an unstable brane to the true vacuum where no perturbative states
present.

The paper is organized as follows. In Section~2 we set up the model
and write down the Friedmann equations. In Section~3 we explore how
the non-local action for the scalar field linearized near a
non-perturbative vacuum can be rewritten in a local form. In
Section~4 we compute cosmological quantities for a general operator
$\Fc$ and an arbitrary background as well as formulate a
perturbative approach for solving the Friedmann equations. In
Section~5 we consider the cosmological dynamics of just the tachyon
in the initially flat space, i.e. $H(t_0)=0$ and show that nothing
can be generated dynamically. In Section~6 we analyze two specific
cosmological backgrounds keeping operator $\Fc$ arbitrary. In the
last Section the obtained results are discussed.


\section{Model set-up}


The action for the tachyon in the Cubic Super SFT~\cite{AMZ,PTY} in
the flat background\footnote{We always use the signature
$(-,+,+,+,\dots)$.} when fields up to zero mass are taken into
account is found to be \cite{NPB,AJK}
\begin{equation}
S_{\Phi}=\frac1{g_o^2}\int d^dx\left(\frac12\Phi\Fc(\Box)\Phi-
\frac{1}{4}\Phi^4\right) \label{CSSFT}
\end{equation}
with
\begin{equation}\label{Dsimple}
\Fc(\omega^2)=\left(\xi^2\omega^2+1\right)e^{-\frac14\omega^2}.
\end{equation}
where $\xi^2\approx 0.9556$ is a constant entirely determined by SFT
and we put $\AP=1$. Scalar field $\Phi$ is the open string tachyon
in question. This action is obviously a non-local one since it
contains an infinite number of derivatives. Further it may be
interesting to generalize the latter expression to bring it to a
$p$-adic string like form by replacing 4 with $p+1$ as follows
\begin{equation}
S=\frac1{g_o^2}\int d^dx\left(\frac12\Phi\Fc(\Box)\Phi-
\frac{1}{p+1}\Phi^{p+1}\right) \label{mainS}
\end{equation}
and probably consider general functions $\Fc$. An equation of motion
for field $\Phi$ reads
\begin{equation}
\Fc(\Box)\Phi=\Phi^{p}. \label{eom_phi}
\end{equation}
Note that $\xi$ does not depend on $p$ or $d$. Also the limit
$p\to1+$ should be taken with care and for this purpose one cannot
fix $\AP$.


Cosmological scenarios built on action (\ref{mainS}) are given by
the following covariantization which accounts a minimal coupling of
the tachyon to the gravity
\begin{equation}
S=\int
d^dx\sqrt{-g}\left(\frac{R}{2\kappa^2}+\frac1{g_o^2}\left(\frac{1}{2}
\Phi\Fc(\Box_g)\Phi-
\frac{1}{p+1}\Phi^{p+1}(x)-\Lambda_\Phi-T\right)\right)\label{mainScosm}
\end{equation}
where
\begin{equation*}
\Box_g=\frac1{\sqrt{-g}}\pd_{\mu}\sqrt{-g}g^{\mu\nu}\pd_{\nu}
\end{equation*}
is the Beltrami-Laplace (BL) operator. Here $g$ is a metric,
$\kappa$ is a gravitational coupling constant $\kappa^2=8\pi
G=\frac1{M_P^2}$ and we choose such units that it is dimensionless,
$T$ encodes perfect fluids which may be considered including the
cosmological constant $\Lambda$. $\Lambda_\Phi$ is also a constant
but we separate it from the cosmological one. It is considered as a
part of a scalar field potential so that in the picture where the
scalar field potential has a non-perturbative minimum $\Lambda_\Phi$
cancels its energy. We define for the sequel $m_p^2\equiv
g_0^2M_P^2$. Though this is obviously not a full theory which may
come form open-closed string interactions it is obvious that such a
minimal coupling gives a starting point to have an insight into the
problem of an open string modes behavior in a curved space-time.

In the present analysis we focus on the four dimensional Universe
with a spatially flat FRW metric which can be written as
\begin{equation}
g_{\mu\nu}=\diag(-1,a^2,a^2,a^2)\label{metric_FRW}
\end{equation}
with $a=a(t)$ being a space homogeneous scale factor. In this
particular case the BL operator is expressed as
\begin{equation}
\Box_g=-\pd_t^2-3H\pd_t+\frac1{a^2}\pd_{x_i}^2\label{BL_FRW}
\end{equation}
where $H\equiv\dot{a}/a$ is the Hubble parameter and the dot denotes
the time derivative. We discuss only a time-dependent scalar field
as well. Thus we can think about the BL operator just as
\begin{equation*}
\BDc=-\pd_t^2-3H\pd_t.
\end{equation*}
If all the fluids in action (\ref{mainScosm}) are coupled one to
each other only through the gravity then equations of motion
following in this case are
\begin{subequations}
\begin{eqnarray}
3m_p^2H^2&=&\rho_\Phi+\sum_i\rho_i,\label{eom_cosm_rho}\\
m_p^2(3H^2+2\dot H)&=&-(p_\Phi+\sum_ip_i),\label{eom_cosm_p}\\
\Fc(\BDc)\Phi&=&\Phi^p\label{eom_cosm_phi},\\
\dot\rho_i+3H(1+w_i)\rho_i&=&0\text{ for any }i.\label{eom_cosm_i}
\end{eqnarray}
\label{eom_cosm}
\end{subequations}
Here $\rho$-s are energies and $p$-s are pressures of fluids and $i$
enumerates perfect fluids coming from the $T$-term in the action.
$\rho_\Phi$ and $p_\Phi$ account $\Lambda_\Phi$. Equations
(\ref{eom_cosm_phi}) and (\ref{eom_cosm_i}) are consequences of the
covariant energy-momentum tensor conservation. So, they are not
independent. Since we take $w_i$ being constants equation
(\ref{eom_cosm_i}) can be easily solved to give
\begin{equation*}
\rho_i=r_i\left(\frac{a_0}{a}\right)^{3(1+w_i)}\text{ and
}p_i=w_i\rho_i.
\end{equation*}
$r_i$ and $a_0$ are constants giving an energy density and the scale
factor at some specific time point $t_0$. One of three remaining
equations (\ref{eom_cosm_rho})-(\ref{eom_cosm_phi}) is not
independent and it will found convenient in the sequel to work with
(\ref{eom_cosm_phi}) and the following equation
\begin{equation}
\frac{\ddot
a}a=-\frac1{6m_p^2}\left(\rho_\Phi+3p_{\Phi}+\sum_i(1+3w_i)\rho_i\right).\label{eom_cosm_ddota}
\end{equation}
Note that $w_\Phi$ which is defined through $p_\Phi=w_\Phi\rho_\Phi$
is not a constant.


\section{Asymptotic tachyon spectroscopy}

Focusing on open string modes we want to emphasize here general
facts about a spectrum of Lagrangian (\ref{mainS}) in an asymptotic
regime. We start specifying two desired properties of operator
$\Fc$:
\begin{itemize}
\item $\Fc(z)$ is an analytic function on a complex plane;
\item $\Fc(z)$ admits Taylor expansion at $z=0$ such that $\Fc(z)=c_nz^n$ where
$c_0=1$ and all $c_n$ are real.
\end{itemize}
These assumptions are very general. Also we restrict ourselves to
have $p>1$ so that there is a non-perturbative vacuum at $\Phi=1$.
In an important case of odd $p$ we also have a symmetry
$\Phi\to-\Phi$ so that $\Phi=-1$ also becomes a vacuum. A potential
is
\begin{equation}
V=-\frac12\Phi^2+\frac1{p+1}\Phi^{p+1}+\Lambda_\Phi. \label{mainV}
\end{equation}
In cubic super SFT one gets $p=3$. The zero value of the potential
in the minimum is assured by choosing
$\Lambda_\Phi=\frac{p-1}{2(p+1)}$.

The picture we have in mind is a rolling tachyon \cite{Sen-g} which
starts rolling from an unstable perturbative vacuum $\Phi=0$ and
approaches a non-perturbative one in an infinite time. Thus an
asymptotic we are going to study is $\Phi=1-\psi$. A linearization
around the true vacuum gives the following action \cite{Arefeva,AK}
\begin{equation}
S=\frac1{g_o^2}\int dx\sqrt{-g}\left(\frac12\psi\Fc(\BDc)\psi- \frac
p2\psi^2\right). \label{mainSas}
\end{equation}
This can be considered as an effective action in the true vacuum of
the SFT. According to the Sen conjectures \cite{Sen-g} we expect
that there should not be open string excitations. This is simply
translated here imposing that operator $\Fc(\BDc)-p$ has no zeros
for finite $\omega^2$ which are eigenvalues of the BL operator. This
is so for approximate operator (\ref{Dsimple}) if again $p>1$.

An approach we use is based on the Weierstrass product method and
has been studied in details in the recent paper by I.Ya.~Aref'eva
and I.V.~Volovich \cite{AV} (also \cite{AJV}). Here we just mention
main steps.

Any eigenfunction of the BL operator with an eigenvalue $\omega^2$
is an eigenfunction of the $\Fc$ operator if $\Fc(\omega^2)$ is well
defined. Here we do not specify rigorously a class of functions on
which we define an action of $\Fc$ just mentioning that these are
smooth complex valued functions on the real axis with all order
derivatives well defined. However, a question of an asymptotic
behavior and an integrability is open. For instance, exponentially
growing modes may be physically important if we consider the model
from the point of view of cosmological perturbations. Thus
\begin{equation*}
\BDc f_\omega=\omega^2
f_\omega~\Rightarrow~\Fc(\BDc)f_\omega=\Fc(\omega^2)f_\omega.
\end{equation*}
However, in general $\Fc(\omega^2)$ may have (infinitely) many
branches thus giving that (infinitely) many functions $f_\omega$ may
correspond to the same eigenvalue of the $\Fc$ operator. In other
words $\Fc$ may have (infinitely) degenerate eigenvalues and this is
a case of operator (\ref{Dsimple}) which has all eigenvalues
infinitely degenerate \cite{AK}.

Provided we can solve the characteristic equation
\begin{equation}\Fc(\omega^2)=p\label{characteristic}\end{equation}
to find all $\omega_k^2$ then it is possible to reformulate
asymptotic action (\ref{mainSas}) as an (infinite) sum of
noninteracting scalar fields as follows
\begin{equation}
S=\frac1{g_o^2}\int
dx\sqrt{-g}\frac12\sum_k\epsilon_k\psi_k(\BDc-\omega_k^2)(\BDc-{\omega_k^2}^*)\psi_k.\label{mainSassum}
\end{equation}
Here $\epsilon_k$ are constants. Recall that equation
(\ref{characteristic}) should not have real roots. Thus all
$\omega^2$ are complex and for each root $\omega_k^2$ exists a
complex conjugate root ${\omega_k^2}^*$ since $\Fc(\omega^2)$ can be
expressed as a polynomial with real coefficients. Equation of motion
looks like
\begin{equation}
(\BDc-\omega_k^2)(\BDc-{\omega_k^2}^*)\psi_k=0\label{eom_assum}
\end{equation}
and has a general four parametric solution and these parameters are
enough to make $\psi_k$ real. Indeed, solution for $\psi_k$ may be
written as
\begin{equation}
\psi_k=\alpha_+\psi_{k+}+\alpha_-\psi_{k-}+\bar\alpha_+\psi_{k+}^{*}+\bar\alpha_-\psi_{k-}^{*}
\label{solution_deltan}
\end{equation}
where $\psi_k^{(i)}$ are two linear independent solutions to the
equation $(\BDc-\omega_k^2)\psi_k=0$. Obviously complex conjugate
functions solve the second order differential equation with a
conjugate $\omega_k^{2*}$ and all the four terms in the latter
expression are linear independent. $\alpha$-s are integration
constants to be adjusted giving real $\psi_k$. Each such a $\psi_k$
solves an equation of motion coming from action (\ref{mainSas}).
Since this equation is linear in $\psi$ then an arbitrary linear
combination of $\psi_k$ does solve it.

So, we have reproduced the spectrum by virtue of an (infinite) sum
of non-interacting scalar fields. A comment is in place here. One
can say that a more simple Lagrangian than (\ref{mainSassum}) may be
formulated where only first power of the BL operator enters the
expression. It is formally possible but since $\omega^2_k$ are
complex one would deal in this case with a non-hermitian Lagrangian
\begin{equation}
S=\frac1{g_o^2}\int dx\sqrt{-g}
\frac12\sum_k\left(\varepsilon_k\psi_k(\BDc-\omega_k^2)\psi_k+\bar\varepsilon_k\bar{\psi}_k(\BDc-{\omega_k^2}^*)\bar{\psi}_k\right).\label{mainSassumc}
\end{equation}
Here $\varepsilon_k$ and $\bar\varepsilon_k$ are constants.
Equations of motion are
\begin{equation}
(\BDc-\omega_k^2)\psi_k=0,\quad(\BDc-{\omega_k^2}^*)\bar\psi_k=0.\label{eom_assumc}
\end{equation}
They have general two parametric solutions
\begin{equation}
\psi_k=\alpha_+\psi_{k+}+\alpha_-\psi_{k-},\quad\bar\psi_k=\bar\alpha_+\bar\psi_{k+}+\bar\alpha_-\bar\psi_{k-}.
\label{solution_deltanc}
\end{equation}
Constants $\varepsilon$ are arbitrary so far. They are fixed by the
Weierstrass product to be $\varepsilon_k=\Fc^\prime(\omega_k^2)$ and
$\bar\varepsilon_k=\Fc^\prime({\omega_k^2}^*)$. In the next Section
we will show explicitly that action (\ref{mainSassumc}) reproduces
correctly an energy-momentum tensor of action (\ref{mainSas}) if
$\psi=\sum_k\psi_k+\sum_k\bar\psi_k$and $\bar\psi_k=\psi_k^*$. Being
possible one should use the latter Lagrangian with care and
solutions for $\psi_k$ will be necessary complex in order to make
full $\psi$ real. It is obvious, that $\bar\psi\sim\psi^*$. However,
a situation is different compared to action (\ref{mainSassum}). A
requirement for solution to equation of motion (\ref{eom_assum}) to
be real is natural and this guaranties a reality of $\psi$. In case
of latter action (\ref{mainSassumc}) the requirement of a reality of
$\psi$ which is $\sum_k\psi_k+\sum_k\bar\psi_k$ is an external one.
Nevertheless, the latter form of the action is found to be useful.
Also we have to say that the above construction becomes more
involved in case of multiple roots\footnote{I would like to thank
S.Yu.~Vernov for pointing this out.} \cite{AV,AJV}.


\section{Cosmology in tachyon vacuum}

In this Section we are going to cosmological scenarios described by
action (\ref{mainScosm}). The main goal is to investigate the
tachyon near its true vacuum.


\subsection{Derivation of cosmological quantities}

The energy-momentum tensor is derived by means of
\begin{equation*}
T_{\alpha\beta}=-\frac2{\sqrt{-g}}\frac{\delta S}{\delta
g^{\alpha\beta}}.
\end{equation*}
In these notations $T_{00}=\rho$ and $T_{ii}=a^2p$. For a general
$\Fc$ the only series expansion in powers of the BL operator may
lead to a result. For action (\ref{mainScosm}) linearized with
$\Phi=1-\psi$ one yields
\begin{eqnarray}
\rho_{\psi}&=&\frac{K+P}2,\quad
p_{\psi}=\frac{K-P}2\label{sumrhop}\\
\text{where }K&=&\sum_{n=1}^\infty
c_n\sum_{l=0}^{n-1}\pd_t\BDc^l\psi\pd_t\BDc^{n-1-l}\psi,\quad
P=\sum_{n=1}^\infty
c_n\sum_{l=0}^{n-1}\BDc^l\psi\BDc^{n-l}\psi\nonumber
\end{eqnarray}
where $c_n$ come from $\Fc(z)=c_nz^n$. One can check that
$\nabla_\mu T^\mu_\nu=0$ on equation of motion (\ref{eom_phi}).
Expression for the state parameter $w_\Phi$ reads
\begin{eqnarray}
w_\Phi&=&\frac{p_\Phi}{\rho_\Phi}=\frac{K-P}{K+P}=-1+\frac{2K}{K+P}.
\label{eom_cosm_wPhisingle}
\end{eqnarray}
The total effective equation of state parameter is given by
\begin{eqnarray}
w&=&\frac{p_\Phi+\sum_ip_i}{\rho_\Phi+\sum_i\rho_i}=\frac{K-P+\sum_iw_i\rho_i}{K+P+\sum_i\rho_i}=
-1-\frac23\frac{\dot H}{H^2}. \label{eom_cosm_wsingle}
\end{eqnarray}

Further manipulations can be performed in the latter expressions
using $\psi=\sum_n\psi_n$ and representation (\ref{solution_deltan})
where all the components are eigenfunctions of the BL operator with
known eigenvalues. For real $\psi$ we should take
\begin{equation*}
\psi=\sum_k\left(\psi_{k+}+\psi_{k+}^{*}+\psi_{k-}+\psi_{k-}^{*}\right)
\end{equation*}
where all integration constants $\alpha$ are included in $\psi$-s
and
\begin{equation*}
\BDc\psi=\sum_k\left(\omega_k^2\psi_{k+}+{\omega_k^2}^*\psi_{k+}^{*}+\omega_k^2\psi_{k-}+{\omega_k^2}^*\psi_{k-}^{*}\right).
\end{equation*}
Surprisingly the sums over $n$ and $l$ can be evaluated in a closed
form to yield the following simple expressions
\begin{equation}
\begin{split}
K&=\sum_k\left(\Fc^\prime(\omega_k^2)\left(\dot\psi_{k+}+\dot\psi_{k-}\right)^2+\Fc^\prime({\omega_k^2}^*)
\left(\dot\psi_{k+}^{*}+\dot\psi_{k-}^*\right)^2\right),\\
P&=\sum_k\left(\omega_k^2\Fc^\prime(\omega_k^2)\left(\psi_{k+}+\psi_{k-}\right)^2+{\omega_k^2}^*\Fc^\prime({\omega_k^2}^*)
\left(\psi_{k+}^{*}+\psi_{k-}^*\right)^2\right).
\end{split}
\label{eom_cosm_KPsummed}
\end{equation}
Here prime denotes a derivative with respect to an argument. The
main achievement at this stage is that all the information can be
extracted by means of solving algebraic equation
(\ref{characteristic}) and constructing of eigenfunctions of the BL
operator. Sum over $k$ is indefinite until a subset of
eigenfunctions of interest is not specified.

There are two remarkable properties worth to mention. First, the
coefficients can be expressed entirely in terms of function $\Fc(z)$
without annoying summations. This is a great simplification and an
opening for a possibility of studying general operators $\Fc$.
Second, there are no mixed terms involving $\psi$-s for different
$k$ as well as $\psi_i\psi_j^*$ combinations. Thus we have shown
explicitly that the energy of scalar field $\Phi$ can be derived
from action (\ref{mainSassumc}) if we put
$\varepsilon_k=\Fc^\prime(\omega_k^2)$ and
$\bar\varepsilon_k=\Fc^\prime({\omega_k^2}^*)$ and take a special
solution $\bar\psi_k=\psi_k^*$.

A mechanism how specific excitations can be selected dynamically is
hidden in initial conditions for field $\Phi$ at zero time and a
correspondence between these conditions and a late time behavior is
not completely clear in a curved background.

We stress that all the above consideration in this Subsection does
not depend on a background and applicable to a general form of the
BL operator.


\subsection{Perturbative approach}

The relevant construction of the previous Section does not rely on a
specific background and is valid for a very general operator $\Fc$
and we want to emphasize the following common points.

First, the spectrum of Lagrangian (\ref{mainSassumc}) does not
depend on a background. Only the form of operator $\Fc$ determines
eigenvalues $\omega_k^2$. Thus, changing a background most probably
does not change the spectrum. There may be, however, very special
backgrounds where some of eigenvalues $\omega_k^2$ may not have
corresponding non-zero eigenfunctions. The latter issue is closely
related to a rigorous definition of a class of functions which
$\psi$ belongs to.

Second, in a wide range of meaningful functions $H(t)$ the spectrum
of the BL operator is the complex plane not restricted in any way
unless we do not restrict our physical states. Thus, changing $\Fc$
without changing a background will shift eigenvalues without major
changes to the physics. Hence changing of a background is the most
significant modification. This in turn changes the BL operator and
its eigenfunctions.

Third, in general knowing eigenfunctions of the BL operator in the
flat space we expect an oscillating behavior of $\psi$ since all
$\omega$-s are complex. Manipulating with $\alpha$-s in
(\ref{solution_deltanc}) we might be able to construct vanishing or
going to wild oscillations solutions. We would not avoid
oscillations completely. There is, however, one specific case when
for some $\omega_n^2$ oscillations in cosmological quantities will
not die or grow. It may be a very interesting situation.

Having the tachyon coupled to a dynamical gravity we have to solve
all the system of cosmological equations (\ref{eom_cosm}). So in
general a background is not fixed and subject to dynamical
equations. Along scalar field $\Phi$ we should find corrections to
the metric, i.e. scale factor $a$ in our case. Zero approximation
for $a$ denoted by $a^{(0)}$ is an expression given by perfect
fluids coming from the $T$ term in action (\ref{mainScosm}) only.
Indeed, $a^{(0)}$ is the scale factor calculated for $\Phi=1$.
$\Lambda_\Phi=\frac{p-1}{2(p+1)}$ which is aimed to cancel a
negative vacuum energy of the scalar field disappears and we are
left with term $T$ only. The latter, however, also may contain a
constant which we name the cosmological constant. Therefore term $T$
forms what we name a ``background'' while interaction with the
scalar field produces ``corrections''. The following perturbative
strategy can be used:
\begin{itemize}
\item We take $a=a^{(0)}$ and solve an equation of motion for the scalar field which is now
(\ref{eom_assumc});
\item We substitute resulting $\Phi=1-\sum_n\psi_n$ in
(\ref{eom_cosm_ddota}) and find correction $a^{(1)}$.
\end{itemize}
Equation (\ref{eom_cosm_ddota}) linearized around $a^{(0)}$ becomes
\begin{equation}
-\ddot
a^{(1)}+v(t)a^{(1)}=\frac{a^{(0)}}{6m_p^2}(\rho_\psi+3p_\psi)=\frac{a^{(0)}}{6m_p^2}(2K-P)
\label{eom_cosm_ddota1}
\end{equation}
where
\begin{equation*}
v(t)=\frac{\ddot
a^{(0)}}{a^{(0)}}+\frac1{2m_p^2}\sum_i(1+w_i)(1+3w_i)r_i\left(\frac{a_0}{a^{(0)}}\right)^{3(1+w_i)}.
\end{equation*}
This procedure can be repeated with found first order corrections to
find next order perturbations. This is beyond of our present
analysis but one comment is very important. In general this may
bring out corrections of order $\psi^2$ indicating thereby that a
linear approximation around the true vacuum is no longer valid and
higher corrections should be accounted in the action.


\section{Tachyon in initially flat space}

We start with the most simple example. Namely, we take $T=0$ meaning
that only the tachyon does present and assume that it is in the
asymptotic regime. Such a situation may be if during the evolution
of the tachyon up to some time a back reaction of the gravity is
negligible. Also for simplicity we take only one mode in $\psi$. In
this case taking $\psi=\alpha+i\beta$, $\omega^2=m^2+ik^2$,
$\Fc^\prime(\omega^2)=x+iy$ and assuming $\bar\psi=\psi^*$ we have
the following action for fields $\alpha$ and $\beta$
\begin{equation}
\begin{split}
S=\frac1{g_o^2}\int
dx\sqrt{-g}&\left(\alpha(x\BDc-xm^2+yk^2)\alpha-\beta(x\BDc-xm^2+yk^2)\beta-\right.\\
&\left.-2\alpha(y\BDc-ym^2-xk^2)\beta\right).
\end{split}
\label{Lempty}
\end{equation}
We see, that for any signs of parameters one normal and one phantom
field present in the system. This is in accord with Ostrogradski
method \cite{ostrogradski,o2} (see \cite{AV} for a discussion on
this point). Note, that only field $\alpha$ is physical one since
$\psi+\psi^*=2\alpha$ is a fluctuation around the minimum of the
tachyon potential. Constants $m^2$ and $k^2$ are certainly real but
not necessarily positive. Moreover, $k^2\neq0$. Thus, the dynamics
of the tachyon near the vacuum is governed effectively by two scalar
fields one of which is a phantom. The action may serve as a toy
model for the tachyon around its vacuum. The energy and pressure
associated with the above Lagrangian are as follows
\begin{equation}
\begin{split}
\rho&=x\dot\alpha^2-x\dot\beta^2+2y\dot\alpha\dot\beta+(xm^2-yk^2)(\alpha^2-\beta^2)-2(ym^2+xk^2)\alpha\beta,\\
p&=x\dot\alpha^2-x\dot\beta^2+2y\dot\alpha\dot\beta-(xm^2-yk^2)(\alpha^2-\beta^2)+2(ym^2+xk^2)\alpha\beta.
\end{split}
\label{rhopempty}
\end{equation}
Friedmann equations for the above action minimally coupled to the
gravity read
\begin{equation}
\begin{split}
3m_p^2H^2&=\rho,\\
m_p^2\dot H&=-x\dot\alpha^2+x\dot\beta^2-2y\dot\alpha\dot\beta.
\end{split}
\label{FriedmannEmpty}
\end{equation}
Plus to this two equations of motion for scalar fields are
\begin{equation}
\begin{split}
(x\BDc-xm^2+yk^2)\alpha-(y\BDc-ym^2-xk^2)\beta&=0,\\
(x\BDc-xm^2+yk^2)\beta+(y\BDc-ym^2-xk^2)\alpha&=0.
\end{split}
\label{EOMEmpty}
\end{equation}
Only three of four above equations are independent. The limit we are
interesting in is $\alpha$ and all its derivatives go to zero, so
the tachyon rests in its vacuum. However, according to equations of
motion (\ref{EOMEmpty}) $\alpha=0$ leads to $\beta=0$ for any real
values of parameters. This is not a desired result demonstrating
that provided $T=0$ the tachyon itself (plus $\Lambda_\Phi$ which
makes its potential energy non-negative) cannot generate any
non-trivial cosmology. A possible scenario that $\alpha$ oscillates
near zero and $\beta$ makes the job of a cosmology generation does
not work. Although there is no analytic prove, the conclusion has
been checked numerically in a wide range of parameters.

However, once we do not require vanishing or bounded field $\alpha$
various possibilities for the cosmological evolution appear. It may
be of great importance to understand whether growing solutions for
$\alpha$ can be physically justified and moreover a mechanism which
suppresses such a growing at some time scale is needed.

Alternatively we can add some ingredient to the dynamics so that the
evolution becomes interesting. Two important examples are developed
in the next Section.


\section{Two more examples}

Below we analyze specific background configurations of perfect
fluids and different possibilities for eigenvalues $\omega_k^2$ and
intensively use obtained formulae (\ref{eom_cosm_KPsummed}) where
the summation has been done for a general operator $\Fc$.


\subsection{Tachyon with cosmological constant}

In paper \cite{AK} operator (\ref{Dsimple}) and $p=3$ situation was
analyzed in a background of the cosmological constant. The first
order corrections were found in this model. Here we relax $p=3$
condition and do not assume any exact form of operator $\Fc$.

Zero approximation for the scale factor is
\begin{equation}
a^{(0)}=a_0e^{H_0t}=a_0e^{\sqrt{\frac{\Lambda}{3m_p^2}}t}.
\label{aCC}
\end{equation}
An equation for eigenfunctions of the BL operator reads
\begin{equation*}
\ddot \psi_{k}+3H_0\dot \psi_k=-\omega_k^2\psi_k
\end{equation*}
and the solution is
\begin{equation}
\begin{split}
\psi_{k+}&=\alpha_{k+}e^{-\omega_{k+}t},\quad
\psi_{k-}=\alpha_{k-}e^{-\omega_{k-}t},\\
\text{where
}\omega_{k\pm}&=\frac32H_0\pm\sqrt{\frac94H_0^2-\omega_k^2},\quad\alpha_{k\pm}\text{
are constants}.
\end{split}
\label{deltaCC}
\end{equation}
Equation (\ref{eom_cosm_ddota1}) becomes
\begin{equation}
-\ddot a^{(1)}+H_0^2a^{(1)}=\frac{e^{H_0t}}{6m_p^2}(2K-P).
\label{a1CC}
\end{equation}
The solution is
\begin{equation}
a^{(1)}=a_+e^{H_0t}+a_-e^{-H_0t}+\bar{a}(t) \label{a1CCsol}
\end{equation}
where the last term is a particular solution to the inhomogeneous
equation. If for any non-zero $\psi_{k+}$ one takes $\psi_{k-}=0$
and vice versa then $K$ and $P$ completely free of mixed terms.
Therefore, since equation (\ref{a1CC}) is linear we can solve it for
only one specific mode and then sum up the answers. Hence, for an
only mode the result for the scalar field is
\begin{equation}
\Phi=1-\alpha e^{-rt}\cos(\nu t+\varphi) \label{phiCC}
\end{equation}
where $\alpha=2|\alpha_{k\pm}|$ and $r$ and $\nu$ are real and
imaginary parts for $\omega_{k\pm}=r_{k\pm}+i\nu_{k\pm}$ with $k$
fixed and a sign chosen. We assume $r\geq0$ so that oscillations do
not grow\footnote{For positive $r$ one easily deduces the results of
\cite{AK}.}. The R.H.S. of equation (\ref{a1CC}) can be represented
as
\begin{equation*}
\frac{a_0e^{(H_0-2r)t}}{6m_p^2}(2\alpha_K\sin(2\nu
t+\varphi_K)-\alpha_P\sin(2\nu t+\varphi_P)).
\end{equation*}
Solution for $\bar a(t)$ can be found to be
\begin{equation}
\begin{split}
\bar
a(t)&=\frac{e^{(H_0-2r)t}}{24m_p^2(r^2+\nu^2)((H_0-r)^2+\nu^2)}\times\\
&\times\left((\nu^2-r^2+H_0r)(2\alpha_K\sin(2\nu
t+\varphi_K)-\alpha_P\sin(2\nu t+\varphi_P))+\right.\\
&\left.+(H_0-2r)\nu (2\alpha_K\cos(2\nu
t+\varphi_K)-\alpha_P\cos(2\nu t+\varphi_P))\right).
\end{split} \label{a1CCih}
\end{equation}
So, correction $a^{(1)}$ is given by (\ref{a1CCsol}) and
(\ref{a1CCih}). Now one can readily get resulting expressions for
$H$ and the state parameter. It is interesting that playing with $r$
and $\nu$ one can observe different behaviors. For instance, for
$r=H_0/2$ oscillations in $a(t)$ will not die despite the fact that
oscillations in $\Phi$ vanish. On the other hand it is impossible to
avoid oscillations completely. These oscillations will be translated
to $w_\Phi$ and effective total state parameter $w$ making possible
periodic transition between phantom and quintessence phases.

Thus we have shown explicitly that a periodic crossing of the
phantom divide may occur as well as a possibility of non-vanishing
oscillations of the cosmological quantities even if the scalar field
tends to its vacuum.


\subsection{Tachyon with perfect fluid $w>-1$}

Title dictates the following equation
\begin{equation}
3m_p^2{H^{(0)}}^2=r\left(\frac{a_0}{a}\right)^{3(1+w)} \label{Hw}
\end{equation}
which can be solved since $H=\dot a/a$. Results for $a^{(0)}(t)$ and
${H^{(0)}}(t)$ are
\begin{equation}
{a^{(0)}}(t)=a_0\left(\frac{t-t_s}{t_0-t_s}\right)^{\frac2{3(1+w)}},\quad
{H^{(0)}}(t)=\frac2{3(1+w)}\frac1{t-t_s}\label{Hwsol}
\end{equation}
where $t_s$ is a constant. An equation for eigenfunctions of the BL
operator reads
\begin{equation*}
\ddot \psi_k+\frac2{(1+w)}\frac1{t-t_s}\dot
\psi_k=-\omega_k^2\psi_k.
\end{equation*}
For arbitrary parameters a solution may be written in terms of
Kummer special functions. However, in the large $t$ limit the
following asymptotic behavior remains
\begin{equation}
\begin{split}
\psi_{k+}&={\alpha_{k+}t^{-\frac1{1+w}}e^{i\omega_kt}},\quad
\psi_{k-}={\alpha_{k-}t^{-\frac1{1+w}}e^{-i\omega_kt}},\\
\dot\psi_{k+}&=\left(i\omega_k-\frac1{(1+w)t}\right)\psi_{k+},\quad
\dot\psi_{k-}=\left(-i\omega_k-\frac1{(1+w)t}\right)\psi_{k-}.
\end{split}
\label{deltaw0}
\end{equation}

To analyze the solution we use $\omega_k=r_k+i\nu_k$. Depending on
the sign of $\nu_k$ either exponentially growing or suppressed
oscillations will be in the dynamics of the scalar field. On top of
this $t^{-\frac1{1+w}}$ factor always persists. Thus even if
$\nu_k=0$ the oscillations will vanish. Result for a single mode in
the scalar field is as follows
\begin{equation}
\Phi=1-\alpha t^{-\frac1{1+w}}e^{-\nu t}\cos(rt+\varphi).
\label{phiw}
\end{equation}
Here $\alpha=2|\alpha_{k\pm}|$, $r$ and $\nu$ are $r_{k}$ and
$\nu_{k}$ for a fixed $k$. Equation (\ref{eom_cosm_ddota1}) reads
\begin{equation}
-\ddot
a^{(1)}+\frac{v_0}{(t-t_s)^2}a^{(1)}=A(t-t_s)^{\frac2{3(1+w)}}(2K-P)
\label{a1w}
\end{equation}
where $v_0=(1+3w)\left(-\frac1{9(1+w)^2}+(1+w)r(t_0-t_s)^2\right)$
and $A=\frac{a_0}{6m_p^2}(t_0-t_s)^{-\frac2{3(1+w)}}$. $v_0$ is not
obviously positive or negative. A solution for $a^{(1)}$ is
\begin{equation}
\begin{split}
&a_+t^{\frac12+\sqrt{\frac14+v_0}}+a_-t^{\frac12-\sqrt{\frac14+v_0}}+\bar{a}(t)\text{
for }v_0>-\frac14,\\
&a_+\sqrt{t}+a_-\sqrt{t}\log t+\bar{a}(t)\text{
for }v_0=-\frac14,\\
&a_+\sqrt{t}\cos\left(\sqrt{-v_0-\frac14}\log
t\right)+a_-\sqrt{t}\sin\left(\sqrt{-v_0-\frac14}\log
t\right)+\bar{a}(t)\text{ for }v_0<-\frac14
\end{split} \label{a1wsol}
\end{equation}
where $t_s$ constant is neglected. The last term is a particular
solution to the inhomogeneous equation. The leading asymptotic in
the R.H.S. of equation (\ref{a1w}) can be represented as
\begin{equation*}
A\alpha_Pt^{-\frac4{3(1+w)}}e^{-2\nu t}\sin(2r t+\varphi_P).
\end{equation*}
Solution for $\bar a(t)$ can be found but the final expression is
very cumbersome even if one fixes $w$. Numeric calculations are
helpful to prove that for any $v_0$ correction $a^{(1)}$ becomes to
be small and vanishing oscillations. The only assumption made is
that we take a vanishing solution for $\psi$. Thus the system looks
like a shaking around a background formed by a perfect fluid.


\section{Discussion and further directions}

To conclude we summarize the main results achieved above.

Non-local action (\ref{CSSFT}) with a general operator $\Fc$ is
analyzed and a local formulation for a linearization near a
non-perturbative vacuum is given by (\ref{mainSassumc}). It does
rely on roots of characteristic equation (\ref{characteristic})
only. The energy and pressure are formulated for a general function
$\Fc(z)$ without specifying its explicit form. Moreover, this
analysis does not depend on a background.

A model example where interaction with gravity is turned on in the
final stage of the tachyon evolution is considered. It is
demonstrated that just a rolling tachyon is not capable to generate
a non-trivial cosmology. A natural question how other components may
enter the action immediately arises.

Two more examples where the scalar field is accompanied by the
cosmological constant or by a perfect fluid are explored in details.
It is shown, in particular, that our scalar field generates a
crossing of the phantom divide in the cosmological constant
background. This crossing is periodic one and moreover, a condition
of non-vanishing oscillations is formulated. The Big Rip singularity
problem is avoided because $w$ exhibits a non-trivial time
dependence and consequently is not a constant less then $-1$.

It would be very interesting to continue the lines of the present
analysis and try to formulate a local action in case of a full
non-linear model. If possible, this will open a way to make use of
numeric methods in analyzing the model. One natural question to be
answered in this way whether the early time dynamics may generate
some cosmic fluids which decouple at the late stage of the tachyon
evolution. This will justify an appearance of an additional term $T$
in the action at large times.

Also it would be interesting to invert the problem and see how
different forms of operator $\Fc$ and the potential affect the
cosmology. This may shade light on a structure of an effective
tachyon action if higher excitations are taken into account in the
SFT.

Another generalization is an inclusion of closed string scalar
fields, the closed string tachyon and dilaton in the analysis to
understand a role of closed string excitations and probably explain
the $T$ term\footnote{See \cite{zw_close} for a discussion on a
closed string tachyon and dilaton condensation.}.


\acknowledgments

The author is grateful to I.Ya.~Aref'eva and S.Yu.~Vernov for useful
comments and discussions. The work is supported in part by Marie
Curie Fellowship MIF1-CT-2005-021982, EU grant MRTN-CT-2004-512194,
RFBR grant 05-01-00758, INTAS grant 03-51-6346 and Russian
President's grant NSh-2052.2003.1.


\end{document}